\DocumentMetadata{}
\documentclass[sigconf,nonacm]{acmart}

\usepackage{listings}
\usepackage{xcolor}
\usepackage{xspace}
\newif\ifdraft

\ifdraft
  \newcommand{\mtnote}[1]{{\textcolor{orange}{ ***Matteo: #1 }}\xspace}
  \newcommand{\miknote}[1]{{\textcolor{red}{ ***Mikhail: #1 }}\xspace}
  \newcommand{\jscnote}[1]{{\textcolor{blue}{ ***John: #1 }}\xspace}
  \newcommand{\jhanote}[1]{{\textcolor{blue}{ ***Shantenu: #1 }}\xspace}

\else
 \newcommand{\mtnote}[1]{}
 \newcommand{\miknote}[1]{}
 \newcommand{\jscnote}[1]{}
 \newcommand{\jhanote}[1]{}
\fi

\begin{document}

\title{Scaling on Frontier: Uncertainty Quantification Workflow Applications
       using ExaWorks to Enable Full System Utilization}

\author{Mikhail Titov}
\authornote{Both authors contributed equally to this work}
\orcid{0000-0003-2357-7382}
\affiliation{%
  \institution{Brookhaven National Laboratory}
  \city{Upton}
  \state{NY}
  \country{USA}
}
\email{titov@bnl.gov}

\author{Robert Carson}
\authornotemark[1]
\orcid{0000-0003-4490-2244}
\affiliation{%
  \institution{Lawrence Livermore National Laboratory}
  \city{Livermore}
  \state{CA}
  \country{USA}
}
\email{carson16@llnl.gov}

\author{Matthew Rolchigo}
\orcid{0000-0003-3439-4207}
\affiliation{%
  \institution{Oak Ridge National Laboratory}
  \city{Oak Ridge}
  \state{TN}
  \country{USA}
}
\email{rolchigomr@ornl.gov}

\author{John Coleman}
\orcid{0000-0002-7261-3143}
\affiliation{%
  \institution{Oak Ridge National Laboratory}
  \city{Oak Ridge}
  \state{TN}
  \country{USA}
}
\email{colemanjs@ornl.gov}

\author{James Belak}
\orcid{0000-0002-1382-3669}
\affiliation{%
  \institution{Lawrence Livermore National Laboratory}
  \city{Livermore}
  \state{CA}
  \country{USA}
}
\email{belak1@llnl.gov}

\author{Matthew Bement}
\orcid{0000-0003-3577-3292}
\affiliation{%
  \institution{Oak Ridge National Laboratory}
  \city{Oak Ridge}
  \state{TN}
  \country{USA}
}
\email{bementmt@ornl.gov}

\author{Daniel Laney}
\orcid{0000-0002-7694-2198}
\affiliation{%
  \institution{Lawrence Livermore National Laboratory}
  \city{Livermore}
  \state{CA}
  \country{USA}
}
\email{laney1@llnl.gov}

\author{Matteo Turilli}
\orcid{0000-0003-0527-1435}
\affiliation{%
  \institution{Rutgers University}
  \city{New Brunswick}
  \state{NJ}
  \country{USA}
}
\affiliation{%
  \institution{Brookhaven National Laboratory}
  \city{Upton}
  \state{NY}
  \country{USA}
}
\email{mturilli@bnl.gov}

\author{Shantenu Jha}
\orcid{0000-0002-5040-026X}
\affiliation{%
  \institution{Rutgers University}
  \city{New Brunswick}
  \state{NJ}
  \country{USA}
}
\affiliation{%
  \institution{Brookhaven National Laboratory}
  \city{Upton}
  \state{NY}
  \country{USA}
}
\email{shantenu@bnl.gov}

\renewcommand{\shortauthors}{Titov and Carson, et al.}


\begin{abstract}
When running at scale, modern scientific workflows require middleware to handle
allocated resources, distribute computing payloads and guarantee a resilient
execution. While individual steps might not require sophisticated control
methods, bringing them together as a whole workflow requires advanced
management mechanisms. In this work, we used RADICAL-EnTK (Ensemble
Toolkit)---one of the SDK components of the ECP ExaWorks project---to implement
and execute the novel Exascale Additive Manufacturing (ExaAM) workflows on up
to 8000 compute nodes of the Frontier supercomputer at the Oak Ridge Leadership
Computing Facility. EnTK allowed us to address challenges such as varying resource
requirements (e.g., heterogeneity, size, and runtime), different execution
environment per workflow, and fault tolerance.
And a native portability feature of the developed EnTK applications allowed
us to adjust these applications for Frontier runs promptly, while ensuring
an expected level of resource utilization (up to 90\%).
\end{abstract}

\ccsdesc[500]{General and reference~Experimentation}
\ccsdesc[500]{Software and its engineering~Real-time systems software}
\ccsdesc[300]{Information systems~Computing platforms}

\keywords{workflow management, runtime system, HPC middleware,
additive manufacturing, Frontier}


\maketitle


\section{Introduction}\label{sec:intro}

The Exascale Additive Manufacturing project (ExaAM)~\cite{exaam_2022}, as part
of the DOE Exascale Computing Project (ECP), has developed a suite of
exascale-ready computational tools to model the
process-to-structure-to-properties (PSP) relationship for
additively manufactured metal components. ExaAM built an uncertainty
quantification (UQ) pipeline to quantify the effect that uncertainty has on
local mechanical responses in processing conditions. The UQ pipeline consists
of 3 main stages, and each stage is represented with a corresponding workflow.

Due to the computational cost of simulating the additive manufacturing process
across time and length scales, GPU-capable machines are required to reduce the
runtime for the whole PSP workflow that ExaAM is targeting. Additionally, 
the number of simulations and node requirements to run a simulation increases when
traversing the later stages. As a part of the UQ pipeline, this ultimately
results in the final stage requiring thousands to tens of thousands of
simulations to be run, with each simulation requiring multiple GPU-capable
nodes. It is this final stage that truly requires not only an exascale
supercomputer in-order to run the full UQ pipeline promptly, but also
an efficient ensemble execution middleware to manage the shear number of 
computational processes being run concurrently.

The UQ pipeline imposes certain requirements on the ensemble management tools:
(i) ability to centralize a streamline of the whole pipeline; (ii) control
different resource requirements (e.g., either having one large batch job for all
workflows or setting a workflow per batch job with the different numbers of
acquired compute nodes and runtime); (iii) support different heterogeneous high
performance computing (HPC) platforms (including different system
architectures); and (iv) fault-tolerance of the tools and executing processes
(i.e., computing tasks that represent batch job steps).

In response to the stated requirements, a corresponding workflow management
toolkit RADICAL-EnTK (Ensemble Toolkit)~\cite{entk_2016}---part of the ExaWorks
Software Development Kit (SDK)~\cite{exaworks_sdk_web}---was evaluated and
chosen. EnTK provides the possibility to: (i) automate runs of different
workflows together, while providing an isolated execution environment per each
workflow; (ii) control the execution state of a workflow and its every task
individually; and (iii) handle the size of a workflow dynamically, e.g., create
a new workflow stages based on the status of previously executed stages.


ExaWorks~\cite{exaworks_2021} -- a DOE ECP project -- integrates independent
middleware to enable the execution of scientific workflows on HPC. ExaWorks
SDK satisfies the workflows requirements of diverse consumers, e.g.,
scientists, facility providers, and developers. It enables teams
to produce portable and scalable applications for a wide range of exascale
workflows. The SDK does not replace the many workflow solutions already
deployed and used by scientists, but rather it provides a robust collection of
community-identified technologies and components tested on Leadership Computing
Facility (LCF) platforms that can be leveraged by users.

ExaWorks SDK comprises a set of workflow management components that feature
clean API designs, enabling them to inter-operate through common software
interfaces. The project is working with the open-source community, application
developers, large computing facilities and HPC platform vendors to create a
sustainable and cross-platform SDK. In particular, the project works with
E4S~\cite{e4s_2023}, which provides from-source builds and containers as well as
robust testing of a broad collection of HPC software packages.

Currently, SDK brings together four seed workflow technologies, specifically
Flux, Parsl, RADICAL-Cybertools, and Swift/T. Other applications and projects,
such as ExaLearn, CANDLE, ExaSky, EXAALT, MARBL, and ExaAM, utilize the ExaWorks
software stack and APIs to implement their workflows.


\section{RADICAL Building Blocks}\label{sec:rct}

RADICAL-Cybertools (RCT)~\cite{rct_2018} are software
building blocks designed to develop efficient and effective tools for scientific
computing. Specifically, RCT allow one to develop scientific applications with up
to 100,000 heterogeneous computing tasks (i.e., a self-contained process, Python
function, or executable, and can be independently executed from other ``tasks'')
and executing them on the largest HPC platforms in
the world at unprecedented scale. RCT support innovative science in multiple
domains, including, but not limited to, drug discovery, climate science,
material science engineering, computational biology and particle physics.


RCT enable writing workflow applications with \textit{task-},
\textit{resource-} and \textit{platform-}level heterogeneity. Each building
block is designed to work as a standalone system or integrated with other
tools from RCT, or third-party software tools. Currently, RCT integrate with
other ExaWorks SDK components,
enabling it to expose diverse functionalities with minimal code editing.
As building blocks and components of an evolving software ecosystem, RCT serve
a vast array of use cases and scientific communities. Due to the ease of
integration within the existing scientific middleware, RCT are well positioned
to support the development of domain-specific frameworks (e.g.,
DeepDriveMD~\cite{deepdrivemd_2019}).


Two of the most commonly used building blocks are RADICAL-Pilot
(RP)~\cite{rp_2022} and EnTK.
RP is a pilot-enabled runtime system that allows users to concurrently execute
up to $10^4$ heterogeneous computing tasks on up to $10^5$ heterogeneous
resources. RP manages concurrent and sequential execution of single/multi
core/GPU/node non/MPI computing tasks on one or more HPC platforms. EnTK is a
workflow engine specifically designed to support the coding and execution of
scientific workflows represented using the PST model. EnTK PST stands for
Pipeline-Stage-Task, where \textit{Pipeline} is a sequence of \textit{Stages},
and each \textit{Stage} is a set of independent computing \textit{Tasks}.
Multiple pipelines can be executed concurrently, while stages, within each
pipeline, are executed sequentially. Grouping tasks into stages represents
dependencies among tasks and enables the concurrent execution of tasks from the
same stage. EnTK utilizes RP to acquire and manage HPC resources, and place and
launch tasks on those resources.


\section{ExaAM Workflows}\label{sec:exaam}

\begin{figure}[htbp]
    \begin{center}
		\includegraphics[width=0.95\columnwidth]{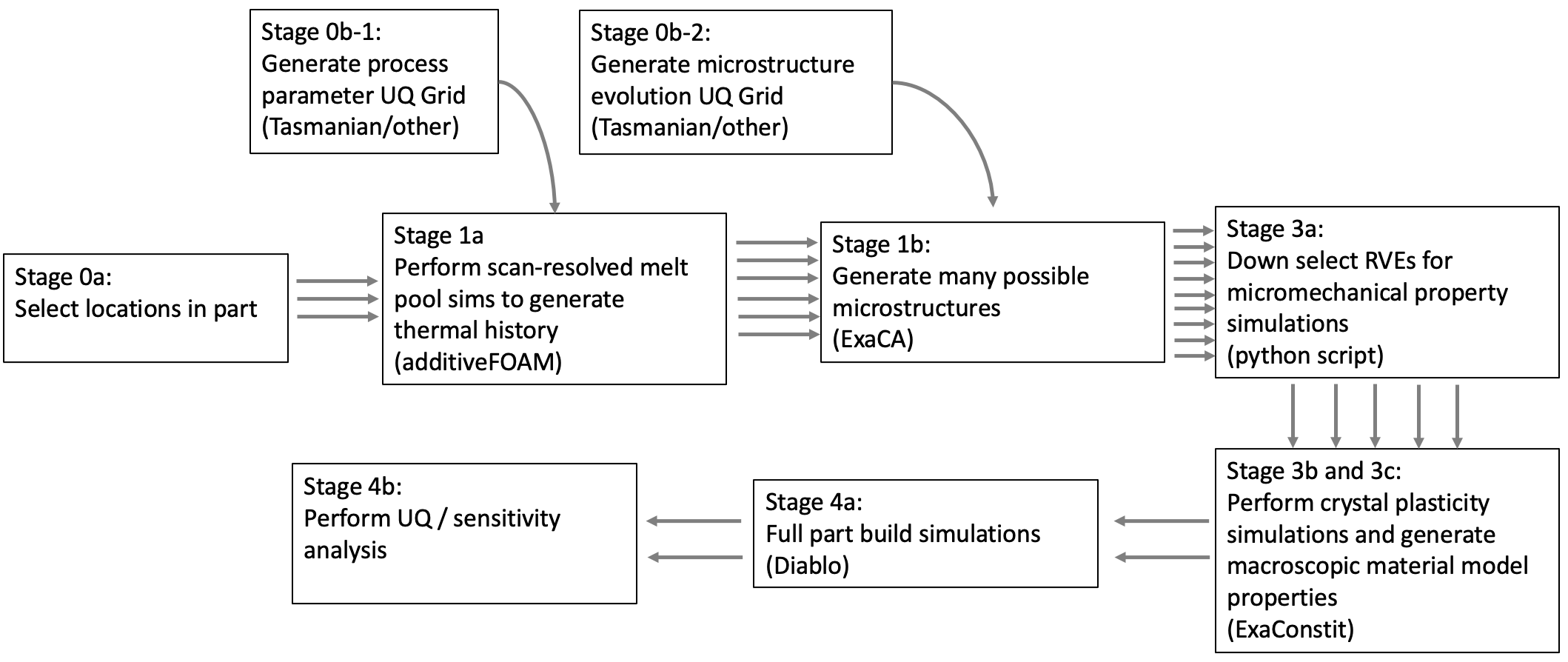}
		\caption{Extended schematic representation of the ExaAM UQ pipeline.}
			\label{fig:uq_workflow}
    \end{center}
\end{figure}

While Fig.~\ref{fig:uq_workflow} shows the full ExaAM UQ pipeline, we will
focus on its three main stages.

\textbf{Stage 0} generates the UQ grid using TASMANIAN~\cite{tasmanian_2013}
and then all the necessary directories and their input files for
\textit{Stage 1}.

\textbf{Stage 1} (melt pool physics + microstructure generation/evolution)
can be represented as two sub-stages, where thermal melt pool simulations are
run in one sub-stage
(\mbox{AdditiveFOAM}~\cite{additivefoam_2023,additivemanufacturing_2023}, an
extension of OpenFOAM for AM processes) and the microstructure generation is
run in a subsequent sub-stage (\mbox{ExaCA}~\cite{exaca_2022}). \textit{Note} that
\mbox{AdditiveFOAM} includes CPU-only computing tasks and requires even and odd
runs to generate all melt pool thermal histories. These runs have an
associated post-processing step to gather all the necessary output files
into a single file for the following sub-stage \mbox{ExaCA}.
\mbox{ExaCA} includes CPU-GPU computing tasks. The microstructure is generated
for the cartesian product between all the thermal pool cases and different
microstructure UQ parameters. Once those runs have completed, a
post-processing step occurs to prepare data for the local property
calculations.

\textbf{Stage 3} (local property calculations) runs all the
ExaConstit~\cite{exaconstit_2019} simulations. It is driven by one
Python script, which reads in all the generated microstructures; coarsens the
microstructures; and generates all the simulation option files and directories
associated with all the different loading directions, temperature cases, and
representative volume elements (RVEs) from \mbox{ExaCA} microstructures. The script
then has a number of built-in job scheduler backends (e.g., Flux, LSF)
available and can run/submit all the jobs in one go. Once all the simulations
are complete, an optimization script then calculates the necessary macroscopic
material model parameters to be used in full part-builds.


We implemented the UQ pipeline as a set of EnTK workflow applications, where
each UQ stage corresponds to the EnTK application (with a single EnTK pipeline)
and consists of one or more EnTK stages. We implemented the
pre-/post-processing operations into dedicated EnTK tasks and grouped them into
EnTK stages. Having a dedicated application per UQ stage allows us to execute
the stages individually or as part of the whole UQ pipeline. The developed code
is located in the project's GitHub repository~\cite{exaam_uq_repo}.


\begin{lstlisting}[
language=Python,
caption=EnTK representation of an \mbox{ExaCA} task on Frontier.,
label={lst:entk-task-example}]
import radical.entk  as re
import radical.pilot as rp

exaca_path = 'ExaCA/build_amd_cp/install/bin'
exaca_task = re.Task({
    'executable' : f'{exaca_path}/ExaCA-Kokkos',
    'arguments'  : ['inputs.json'],
    'pre_exec'   : [
        'module load craype-accel-amd-gfx90a',
        'module load rocm/5.4.0',
        'export CRAYPE_LINK_TYPE=dynamic',
        'export MPIR_CVAR_GPU_EAGER_DEVICE_MEM=0',
        'export MPICH_GPU_SUPPORT_ENABLED=1'
    ],
    'cpu_reqs'   : {'cpu_processes'   : 8,
                    'cpu_threads'     : 7,
                    'cpu_thread_type' : rp.OpenMP},
    'gpu_reqs'   : {'gpu_processes'   : 1}
})
\end{lstlisting}

UQ Stage 1 is transformed into the EnTK application with the following EnTK
stages: AdditiveFOAM's pre-processing and AdditiveFOAM, ExaCA and
ExaCA-Analysis. The various melt pool cases (AdditiveFOAM) and microstructure
generation cases (ExaCA) are represented as single tasks within each of their
corresponding EnTK stages. An example of an \mbox{ExaCA} task implemented with EnTK
is shown in Listing~\ref{lst:entk-task-example}. All of the logic needed to
drive Stage 1 is implemented in one script, which acts as a standalone EnTK
application. This application handles failed tasks by re-submitting them as
part of the consecutive batch job (i.e., the next EnTK run). This automated
process helps to deal with hardware failures to run collected failed tasks
using a new job allocation. During re-submission of failed tasks, the execution
order is preserved according to the order of the original EnTK stages.


UQ Stage 3 integrates a corresponding EnTK application to leverage all the job
ensembles (i.e., set of simulations per batch job).
Additional logic has been added so that each ensemble respects Frontier's job
scheduling policy in terms of walltime limits per amount of requested compute
nodes. Each simulation is represented as a single EnTK task. Task failures
are handled following the same approach as for UQ Stage 1 (re-submitted job size
is smaller and correlates to the number of failed tasks).


Using EnTK (which supports multiple job schedulers such as Flux, LSF, Slurm,
PBSPro, or Cobalt) allowed us to abandon the manual creation and management of
batch scripts in favor of having a single ensemble manager EnTK to handle
everything in one large job or subsequent smaller jobs submissions. We also
introduced fault-tolerance for task execution level, which improved the
efficiency of each EnTK application and UQ pipeline as a whole.


\section{Frontier run}\label{sec:frontier}

RP as a runtime system underneath any EnTK application guarantees portability
of resource management capabilities among HPC platforms with different
architectures. We used Crusher (an early-access testbed platform for the
Frontier system) to evaluate the upcoming exascale system's architecture,
and to pre-configure and adjust RP components. Since every new platform has a
specific configuration for their job scheduler, the runtime system should be
configured accordingly, e.g., number of available cores per node, number of
cores reserved for system processes, configurable options for multiple
concurrent executions per node, etc.

Developed EnTK applications are easily reconfigured for each platform via its
resource configuration and corresponding execution environment setup for every
task type (e.g., see attribute \texttt{pre\_exec} for the EnTK task in
Listing~\ref{lst:entk-task-example}). These applications have been tested on
multiple platforms at OLCF with different job schedulers: Summit (LSF), Crusher
(Slurm), and Frontier (Slurm).

Early runs on Summit and Crusher utilized up to 10 compute nodes for several
hours, and were used to verify the correctness and stability of the execution
process before targeting Frontier. With the scale-up on Frontier, resource
utilization is the following:
\begin{itemize}
  \item
  AdditiveFOAM workflow utilized 40 compute nodes for 2 hours (every
  task requires 4 nodes with 56 cores per node);
  \item
  ExaCA workflow utilized 125 compute nodes for 4 hours (every task
  requires 1 node and utilizes 8 MPI ranks with \texttt{7CPUs-1GPU}
  decomposition);
  \item
  ExaConstit workflow utilized 8000 compute nodes ($85\%$ of Frontier's nodes)
  for up to 3.3 hours to run and orchestrate 7875 tasks (every task requires
  8 nodes with 8 MPI ranks per node with the typical \texttt{7CPUs-1GPU}
  decomposition, and runtime $\sim$10-25 min).
\end{itemize}


\begin{figure}[htbp]
    \begin{center}
		\includegraphics[width=0.95\columnwidth]{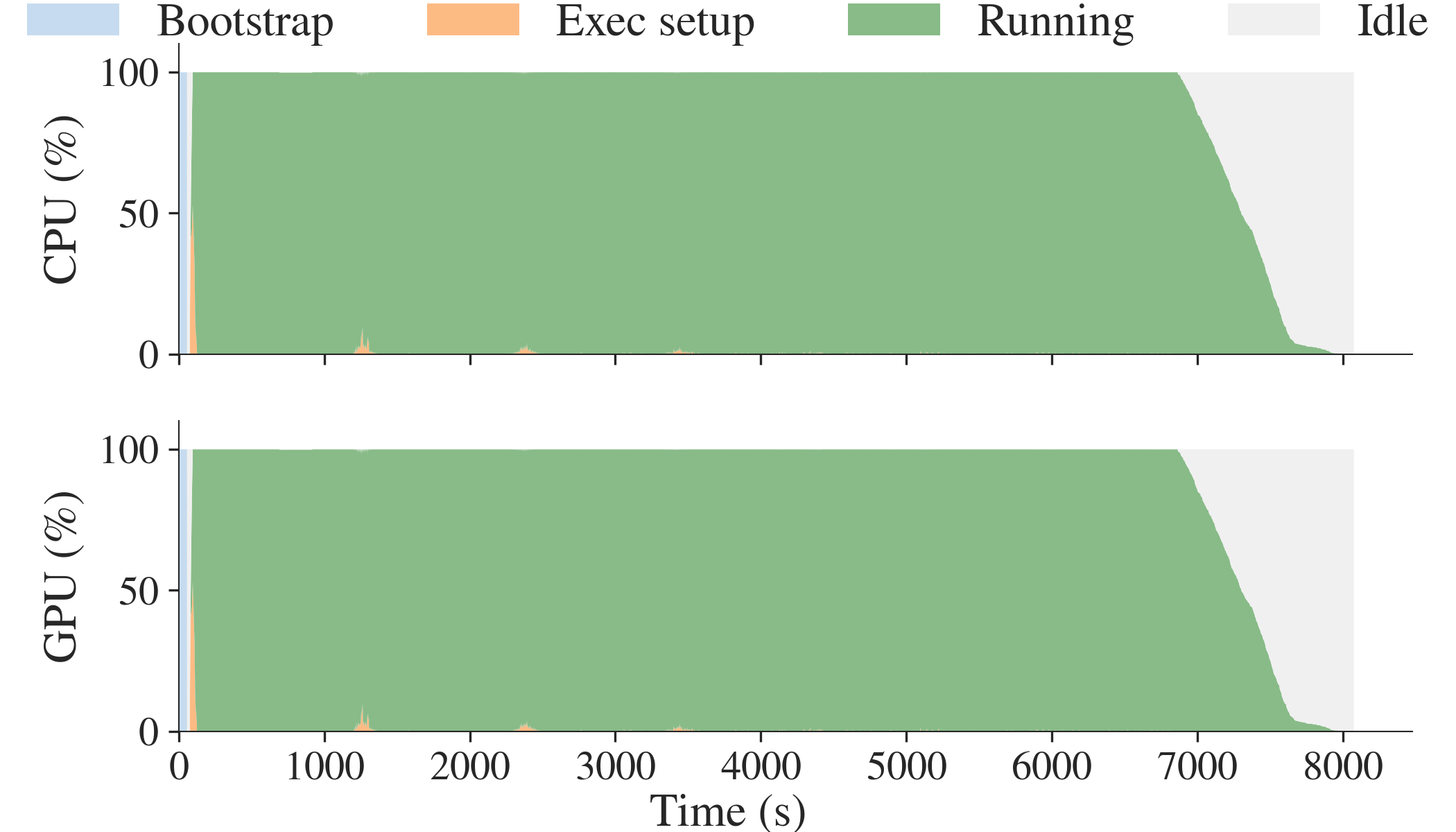}
		\caption{Resource utilization by the EnTK application (UQ Stage 3):
                 100\% corresponds to 448,000 CPU cores (not considering 8 cores
                 per node reserved for system processes) and 64,000 GPUs.}
			\label{fig:entk_job_ru_stack}
    \end{center}
\end{figure}

\begin{figure}[htbp]
    \begin{center}
		\includegraphics[width=0.98\columnwidth]{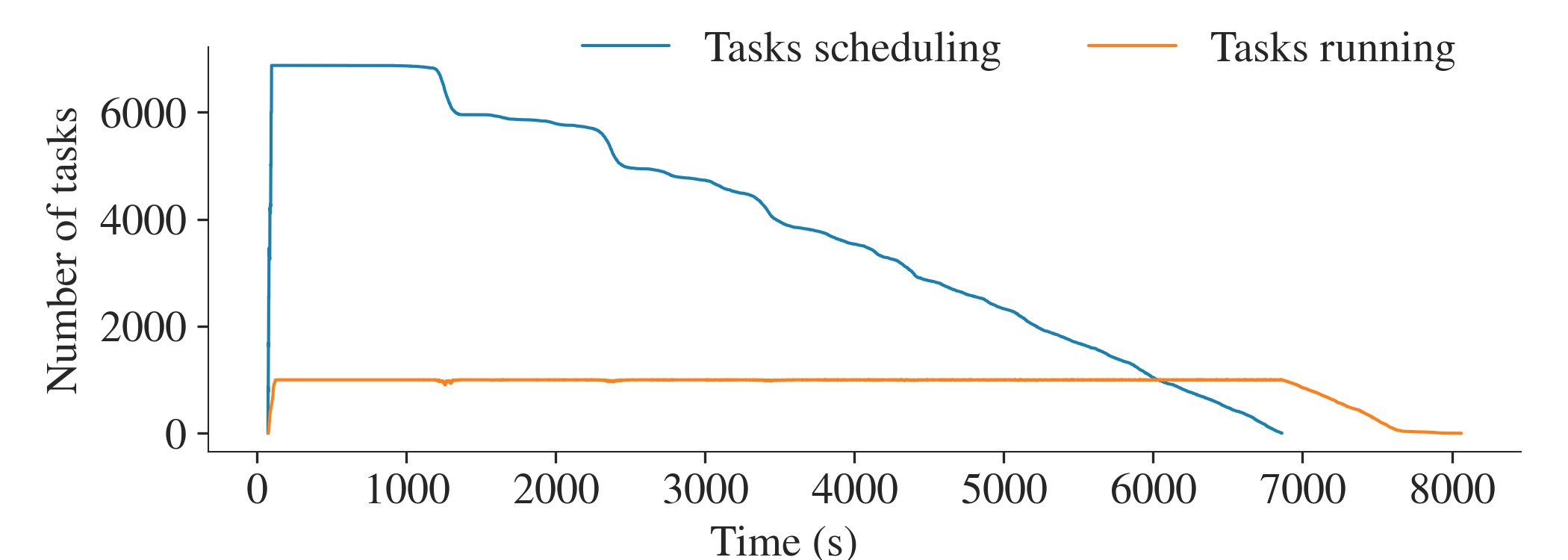}
		\caption{Concurrency of 7875 EnTK tasks (UQ Stage 3) in scheduling
                 and running (execution) states.}
			\label{fig:entk_job_conc}
    \end{center}
\end{figure}

``Full'' scale run for UQ Stage 3 constantly utilized most of the available
resources (total resource utilization is $90\%$) with a minimal overhead (OVH)
of the EnTK application (i.e., bootstrapping EnTK components).
Fig.~\ref{fig:entk_job_ru_stack} shows that OVH (light blue color) is just
85s, while the total execution time of all simulations (TTX) is 7989s (the job
runtime is 8074s). ExaAM workflows implemented with EnTK reached a scheduling
throughput of 269 tasks/s, launching 51 tasks/s. Those rates are part of
Fig.~\ref{fig:entk_job_conc} (initial slopes of blue and orange lines), which
also shows the number of tasks executing concurrently (orange color) as well as
the number of tasks pending to be launched (blue color).

We registered only 10 task failures across the UQ Stage 3 run. Two tasks
failed on the very last simulation step due to too large of a time step for the
specific loading condition and RVE, but they were still far enough out for the
purpose of constructing the macroscopic material model parameters. The other
eight tasks failed due to a single node failure and ran successfully once
automatically re-submitted on Frontier by EnTK.


\section{Summary}\label{sec:summary}

In this paper, we have presented the novel ExaAM UQ pipeline and its
implementation with RADICAL-EnTK. Running simulations for the additive
manufacturing process requires an exascale heterogeneous machine (for CPU-GPU
intensive computations), as well as an efficient ensemble manager to automate
the execution of the whole UQ pipeline and manage varying resource requirements,
different execution environments while offering fault-tolerance. Selecting EnTK
as a workflow engine ensures running the UQ pipeline efficiently and
effectively. Once implemented in EnTK, these applications are portable across
multiple LCF platforms. We tested them on OLCF Summit, Crusher and Frontier.
For runs on Frontier, we progressively increased scale until executing the
\mbox{ExaConstit} workflow with 7875 tasks on 8000 compute nodes ($85\%$ of
Frontier's nodes). We achieved a total resource utilization of $90\%$. We
continue to study the scaling challenge, while increasing the number of
concurrent simulations and conducting experiments with different job
schedulers (e.g., Flux, in case of limitations with Slurm and LSF).



\begin{acks}
This research used resources of OLCF at the Oak Ridge National Laboratory,
which is supported by the Office of Science of the U.S. Department of Energy
under Contract No. DE-AC05-00OR22725. This research was supported by the
Exascale Computing Project (17-SC-20-SC), a collaborative effort of the U.S.
DOE Office of Science and the NNSA, by the ECP ExaWorks project under DOE
Contract No. DE-SC0012704, and under the auspices of the U.S. DOE by
Lawrence Livermore National Laboratory under Contract No. DE-AC52-07NA27344.
We also acknowledge DOE INCITE awards for allocations on Summit, Crusher,
Frontier.
\end{acks}


\bibliographystyle{ACM-Reference-Format}
\bibliography{references}

\end{document}